# Polarization control via artificial optical nonlinearity in dielectric metasurfaces

*Fuyong Yue[1,†,§], Giacomo Balistreri[1,†], Nicola Montaut[1], Fabrizio Riminucci[2,3], Andrea Toma[4], Riccardo Piccoli[1,#], Stefano Cabrini[3], Roberto Morandotti[1]\*, Luca Razzari[1]\**

[1]Institut National de la Recherche Scientifique, Centre Énergie Matériaux Télécommunications (INRS-EMT), 1650 Boulevard Lionel-Boulet, Varennes, Québec J3X 1P7, Canada
[2]Dipartimento di Fisica, Università del Salento, Strada Provinciale Lecce-Monteroni, Campus Ecotekne, Lecce 73100, Italy
[3]Molecular Foundry, Lawrence Berkeley National Laboratory, One Cyclotron Road, Berkeley, California 94720, USA
[4]Istituto Italiano di Tecnologia, Via Morego 30, Genova 16163, Italy

\*E-mail: roberto.morandotti@inrs.ca
\*E-mail: luca.razzari@inrs.ca

[†] These authors contributed equally to this work.
[§] Present address: School of Microelectronics, University of Science and Technology of China, Hefei, Anhui 230026, China
[#] Present address: Department of Molecular Sciences and Nanosystems, Ca' Foscari University of Venice, via Torino 155, 30172, Venice, Italy

**Abstract**
Nonlinear optical phenomena are generally governed by geometry in matter systems, as they depend on the spatial arrangement of atoms within materials or molecules. Metasurfaces, through precisely designed geometries on a subwavelength scale, allow tailoring the optical response of a material far beyond its natural properties. Therefore, metasurfaces are highly appealing to enable the engineering of nonlinear optical interactions at an unprecedented level. Current studies on nonlinear metasurfaces predominantly focus on the phase control of the generated light. Nonetheless, investigating the tensorial nature of the nonlinearity of metasurfaces and its effect on the polarization of the generated light is critical to fully unlock a range of applications, such as nonlinear vector beam generation and nonlinear polarization imaging. Here, we study the artificial optical nonlinearity of a dielectric metasurface originating from its meta-atom symmetry and describe the third-order nonlinear behavior by considering the polarization degree of freedom. We establish an effective nonlinear medium model that serves as a design toolbox for developing amorphous silicon-based geometric metasurfaces with customizable features in third harmonic generation. We further extract quantitative values of the artificial nonlinear susceptibility tensor elements related to the investigated nonlinear process and geometry. The implemented functional devices demonstrate the versatility of dielectric metasurfaces in shaping the emitted light in terms of amplitude, phase, and polarization, for the precise engineering of novel nonlinear architectures targeting applications in nonlinear imaging and complex light generation.

**Keywords:** Third harmonic generation; Nonlinear geometric phase; Dielectric metasurfaces; Polarization engineering; Artificial nonlinear susceptibility; Nonlinear polarization gratings

**Introduction**
Nonlinear optics deals with the study of interactions between intense light fields and optical media, such as crystals and other solids [1, 2], as well as inorganic and organic molecules [3], for achieving desirable functionalities, e.g., optical frequency conversion [1, 4], terahertz generation [5], and stimulated Raman scattering [6]. This enables a wide range of applications in high-resolution optical microscopy [7, 8], spectroscopy and sensing [6, 9, 10], and entangled photon-pair generation [11], to name a few. At the core of these nonlinear effects are the microscopic properties of the matter system (i.e., the arrangement of atoms and molecules, the chemical composition, as well as the lattice symmetries in crystals [12]), which generally determine the tensorial response of its nonlinearity and thus the intrinsic characteristics of the nonlinear interaction. Over the last decade, plasmonic metasurfaces have been employed in nonlinear parametric processes to tailor the emitted light at a subwavelength scale without being constrained by intrinsic material properties and stringent phase-matching requirements [13-15]. Depending on the target application, nonlinear plasmonic metasurfaces are designed through the effective shaping of their individual elements or meta-atoms [16, 17], as well as their lattice distribution (i.e., their arrangement on a substrate) [18], providing an ultra-compact framework for multifunctional optical devices [13, 19-22]. Plasmonic metasurfaces also grant access to continuous phase control of the local nonlinearity for harmonic field generation, which has been demonstrated via the interaction of a circularly polarized fundamental beam with rotated meta-atoms exhibiting certain rotational symmetries [23]. These capabilities have enabled a number of nonlinear applications, e.g., spin-controlled nonlinear holograms [24] and the generation of high-order orbital angular momentum states [25].

In contrast to plasmonic architectures, high-index dielectric metasurfaces exhibit high damage thresholds, can confine the electromagnetic field inside the meta-atoms (rather than over their surface), and feature negligible linear absorption, thus representing an effective alternative platform for the development of miniaturized nonlinear optical devices [26-28].

Nonlinear dielectric metasurfaces composed of silicon meta-atoms with various rotational symmetries have been studied for nonlinear wavefront control during four-wave mixing processes, such as third-harmonic generation (THG) [29, 30]. Despite the extensive literature on these metasurfaces, a comprehensive investigation of the tensorial artificial nonlinearity arising in such systems and its effect on polarization control is still lacking. Indeed, most studies focus on the phase manipulation of the nonlinearly generated light, disregarding valuable information on its full polarization characteristics. Third harmonic (TH) light generated from rotated nanostructures featuring a two-fold rotational symmetry has been reported to consist of both right- and left-circular polarization components, with distinct spin-dependent nonlinear geometric phases, when excited with either a left- or a right-circular polarized fundamental beam [29]. These metasurfaces have been used for complex encoding schemes by mapping phase profiles as a function of the input and output circular polarization, unlocking nonlinear imaging technologies through polarization multiplexing [30]. A quantitative and complete evaluation of the nonlinear characteristics of such metasurfaces is now essential for their effective engineering towards advanced applications encompassing nonlinear imaging, spectroscopy, and nonclassical light generation.

In this work, we present an analytical description of the artificial third-order nonlinearity of metasurfaces originating from their meta-atom symmetry and provide a comprehensive model to describe the physical process of THG. Emphasis is placed on the polarization of the TH light from rotated meta-atoms by considering both the arbitrary polarization state of the fundamental beam and the geometry-induced artificial optical nonlinearity, delivering a convenient design toolbox for nonlinear dielectric metasurfaces. To verify this model, we experimentally investigate the THG characteristics of nonlinear dielectric metasurfaces featuring cuboid-shaped amorphous silicon (a-Si) meta-atoms. A plain metasurface composed of unrotated meta-atoms is designed and fabricated to quantify the tensor elements associated with the geometry-induced artificial nonlinearity. We then apply this knowledge to realize a nonlinear polarization metagrating that can generate TH light with orthogonal linear polarizations in the

zero- and first- diffraction orders. Moreover, we show that the polarization response of multiple diffraction orders of TH light generated by a nonlinear gradient metasurface can be properly accounted for by considering the tensor elements retrieved from the plain metasurface, which further validates our methodology. These and other nonlinear metadevices that can be designed exploiting the presented model have the potential to enable a range of new applications in nonlinear polarimetry [31] and interferometry [32], up-conversion imaging [33, 34], optical encryption [35], as well as the generation of structured light [36] and correlated/entangled photon pairs [37-39].

**Theory**

To obtain a general description of the nonlinear optical interaction in the designed metasurfaces, the meta-atom geometry is treated as a unit cell corresponding to a certain crystal class. In this work, as a proof of concept, we consider cuboid-shaped a-Si meta-atom blocks. Such geometry can be associated with the orthorhombic crystal class (*mmm* in the Hermann-Mauguin notation), since it comprises three mirror planes (m) and three twofold rotation axes ($A_2$) that are perpendicular to each mirror plane, see Fig. 1(a). In the Schoenflies notation, this structure falls within the $C_2$ class, showing a twofold in-plane rotational symmetry [23, 29].

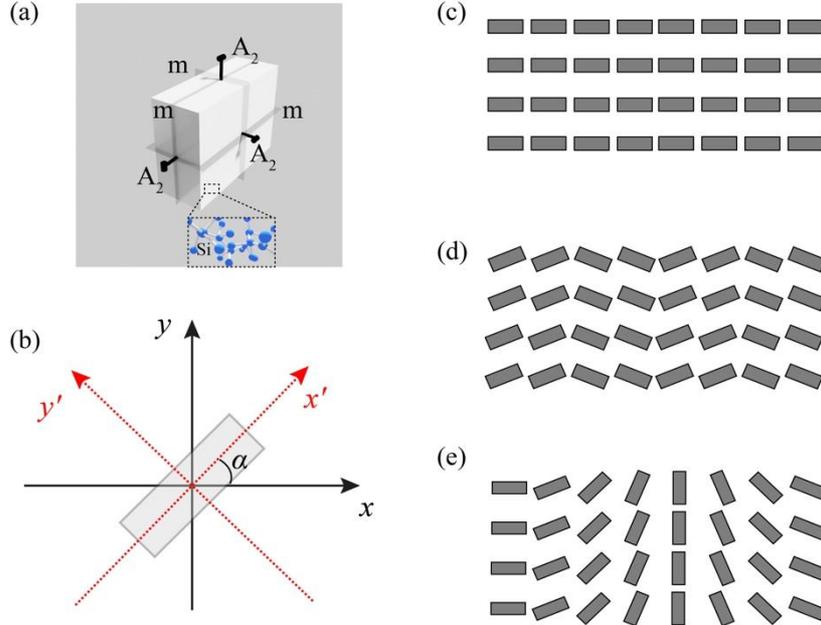

**Figure 1**. (a) Symmetries of an a-Si cuboid block (meta-atom), where m and $A_2$ represent its mirror planes and twofold rotation axes, respectively. The inset depicts the a-Si atomic arrangement. (b) A meta-atom is transformed from the normal coordinate notation (*x*, *y*) to the local coordinates ($x', y'$) via counterclockwise rotation by an angle $\alpha$. Schematic of (c) a plain metasurface, (d) a polarization metagrating, and (e) a gradient metasurface.

We focus on the THG process, which yields the TH frequency of light, $3\omega$, when a fundamental beam with frequency $\omega$ is incident onto the metasurface. The nonlinear polarization density, $\tilde{P}$, is proportional to the third-order susceptibility matrix, $M^{(3)}$ (hereafter expressed without the superscript), associated to the *mmm* orthorhombic crystal class [40],

$$M = \begin{pmatrix} \chi_{11} & 0 & 0 & 0 & 0 & \chi_{16} & 0 & \chi_{18} & 0 & 0 \\ 0 & \chi_{22} & 0 & \chi_{24} & 0 & 0 & 0 & 0 & \chi_{29} & 0 \\ 0 & 0 & \chi_{33} & 0 & \chi_{35} & 0 & \chi_{37} & 0 & 0 & 0 \end{pmatrix}.$$

Here, the compact form $\chi_{lm}$ is used to represent the components of the fourth-rank tensor, where *l* is 1, 2, 3, corresponding to the *x*, *y*, *z* polarization components of the TH light, respectively, and *m* is 1, 2, …, 10, corresponding to the combinations of polarization components of the fundamental light given by *xxx*, *yyy*, *zzz*, *yzz*, *yyz*, *xzz*, *xxz*, *xyy*, *xxy*, *xyz*, respectively. Note that, $M$ is sensitive to the rotation angle of the meta-atoms, $\alpha$ (defined in the counter-clockwise direction in the $x$-$y$ plane, as shown in Fig. 1(b)), which can be accounted for in our theoretical estimations. To simplify our analysis, we consider the case where the incident light is polarized in the $x$-$y$ plane, propagates in the $z$-direction (i.e., $E_z = 0$), and experiences negligible resonance effects within the meta-atoms, similarly to what done in Ref. [29]. We assume that the polarization of the fundamental light remains unchanged within the meta-atoms. The nonlinear susceptibility tensor matrix can thus be reduced to:
$$M = \begin{pmatrix} \chi_{11} & 0 & \chi_{18} & 0 \\ 0 & \chi_{22} & 0 & \chi_{29} \end{pmatrix}.$$

We consider the general scenario in which an arbitrarily polarized fundamental field, described by the Jones vector $\tilde{A} = A_0 \begin{bmatrix} \cos\theta \\ e^{i\delta}\sin\theta \end{bmatrix} = \psi_R |R\rangle + \psi_L |L\rangle$, interacts with rotated meta-atoms (see Fig. 1(b)). Here, $A_0$ is the Jones vector amplitude, $\theta$ denotes the polarization angle, $\delta = \phi_x - \phi_y$ is the phase difference between the $x$- and $y$-polarization components, $|R\rangle$ and $|L\rangle$ are the normalized right- and left-handed circular polarization states (RCP and LCP), respectively, defined as $|R\rangle = \frac{1}{\sqrt{2}} \begin{bmatrix} 1 \\ -i \end{bmatrix}$ and $|L\rangle = \frac{1}{\sqrt{2}} \begin{bmatrix} 1 \\ i \end{bmatrix}$, $\psi_R$ and $\psi_L$ are the corresponding complex amplitudes given by $\psi_R = \frac{A_0}{\sqrt{2}}(\cos\theta + ie^{i\delta}\sin\theta)$ and $\psi_L = \frac{A_0}{\sqrt{2}}(\cos\theta - ie^{i\delta}\sin\theta)$.

The polarization properties of the TH light generated by the rotated meta-atoms are estimated by analyzing the process using a circular polarization basis [23]. The polarization state of the fundamental light is first projected from the normal $(x, y)$ into the local $(x', y')$ coordinates, where the $x'$ and $y'$ axes are along the long and short axes of the rotated meta-atoms, respectively (see Fig. 1(b)). In particular, the Jones vector of the input light under the local coordinates, $\tilde{A}'$, is obtained from the product of the rotation matrix $J(\alpha) = \begin{bmatrix} \cos\alpha & -\sin\alpha \\ \sin\alpha & \cos\alpha \end{bmatrix}$ and $\tilde{A}$. The THG nonlinear polarization density amplitude under the local coordinates, $\tilde{P}'_{3\omega}$, can then be calculated from standard nonlinear optical equations for the THG process [40, 41]. Lastly, the nonlinear polarization density amplitude is transformed back into the normal coordinates $(x, y)$ by multiplying the rotation matrix $J(-\alpha) = \begin{bmatrix} \cos\alpha & \sin\alpha \\ -\sin\alpha & \cos\alpha \end{bmatrix}$ and $\tilde{P}'_{3\omega}$. This gives:

$$\tilde{P}_{3\omega} = (a_1(\psi_R^3 e^{4i\alpha}|L\rangle + \psi_L^3 e^{-4i\alpha}|R\rangle) + a_2(\psi_R^3 e^{2i\alpha}|R\rangle + \psi_L^3 e^{-2i\alpha}|L\rangle) + a_3(\psi_R^2 \psi_L e^{2i\alpha}|L\rangle + \psi_R \psi_L^2 e^{-2i\alpha}|R\rangle) + a_4(\psi_R^2 \psi_L |R\rangle + \psi_R \psi_L^2 |L\rangle))) \quad (1)$$

where
$$\begin{cases} a_1 = \varepsilon_0 \frac{\chi_{11} - 3\chi_{18} + \chi_{22} - 3\chi_{29}}{32} \\ a_2 = \varepsilon_0 \frac{\chi_{11} - 3\chi_{18} - \chi_{22} + 3\chi_{29}}{32} \\ a_3 = \varepsilon_0 \frac{3\chi_{11} + 3\chi_{18} - 3\chi_{22} - 3\chi_{29}}{32} \\ a_4 = \varepsilon_0 \frac{3\chi_{11} + 3\chi_{18} + 3\chi_{22} + 3\chi_{29}}{32} \end{cases} \quad (2)$$

A detailed derivation is presented in Section 1 of the **Supplementary Information**. The resulting polarization density directly acts as a source of TH radiation [7]. Equation (1) thus provides a complete picture of the polarization and phase properties of the generated TH field based on the polarization of the fundamental light and the rotation angle of the meta-atoms.

When an RCP input field illuminates the meta-atoms, the corresponding Jones vector can be represented as $\tilde{A} = A_0 |R\rangle$, i.e., $\psi_R = A_0, \psi_L = 0$. By substituting these values into Eq. (1), we get:
$$\tilde{P}_{3\omega, RCP_{in}} = A_0^3 (a_1 e^{4i\alpha}|L\rangle + a_2 e^{2i\alpha}|R\rangle). \quad (3)$$

The THG nonlinear polarization density (and, as a consequence, the generated TH field) consists of two terms: the first has the opposite handedness of circular polarization (when compared to the fundamental polarization state), LCP, with a geometric phase of $e^{4i\alpha}$, and the second has the same handedness of circular polarization, RCP, with a geometric phase of $e^{2i\alpha}$. Similarly, when an LCP input is used, $\tilde{A} = A_0|L\rangle$, i.e., $\psi_R = 0, \psi_L = A_0$, the nonlinear polarization density amplitude becomes:

$$\tilde{P}_{3\omega,LCP_{in}} = A_0^3(a_1 e^{-4i\alpha}|R\rangle + a_2 e^{-2i\alpha}|L\rangle). \quad (4)$$

The resulting THG nonlinear polarization density shows similar features, with the first term exhibiting the opposite circular polarization, RCP, with a geometric phase of $e^{-4i\alpha}$, and the second term having the same circular polarization of the input, LCP, with a geometric phase of $e^{-2i\alpha}$. These predictions match well with previous results reported in Refs. [23, 29]. Beyond this, note that Eqs. (2)-(4) can quantitatively describe the amplitudes of the circular polarization components of the emitted TH light.

When the meta-atoms are illuminated with a linearly polarized input (i.e., $\delta = 0$) with polarization angle $\theta$, the nonlinear polarization density amplitude takes the form:

$$\tilde{P}_{3\omega,LP_{in}} = \frac{A_0^3}{2}\left(a_1\begin{bmatrix}\cos(3\theta + 4\alpha)\\ -\sin(3\theta + 4\alpha)\end{bmatrix} + a_2\begin{bmatrix}\cos(3\theta + 2\alpha)\\ \sin(3\theta + 2\alpha)\end{bmatrix} + a_3\begin{bmatrix}\cos(\theta + 2\alpha)\\ -\sin(\theta + 2\alpha)\end{bmatrix} + a_4\begin{bmatrix}\cos\theta\\ \sin\theta\end{bmatrix}\right). \quad (5)$$

The resulting TH nonlinear polarization density consists of four terms, where the first three terms depend on both the polarization angle of the fundamental light, $\theta$, and the rotation angle of the meta-atoms, $\alpha$, while the last term depends only on $\theta$ and maintains the input polarization state.

This theoretical framework served as a toolbox for designing nonlinear metasurfaces with customizable polarization properties, as elaborated below. Three kinds of metasurfaces were designed to validate the developed model and extract the effective susceptibility tensor matrix associated with the geometry-induced third-order nonlinear process.

### Device design

*Plain metasurface:* The plain metasurface comprises unrotated ($\alpha = 0°$) meta-atoms with their long axis aligned along the *x*- axis (see Fig. 1(c)). For a linearly polarized input, the output TH electric field and polarization state can be determined based on the input polarization angle and the susceptibility tensor values. As evident from Eq. (5), the output TH light remains linearly polarized, since the associated nonlinear polarization density amplitude can be written as

$$\tilde{P}_{3\omega,LP_{in}} = P_{TH}\begin{bmatrix}\cos(\theta_{TH})\\ \sin(\theta_{TH})\end{bmatrix} \quad (6)$$

, with $P_{TH}(\theta) =$
$\frac{\varepsilon_0 A_0^3}{32}\sqrt{4\cos(\theta)^2(\chi_{11} + 3\chi_{18} + (\chi_{11} - 3\chi_{18})\cos(2\theta))^2 + (-3(\chi_{22} + \chi_{29})\sin(\theta) + (\chi_{22} - 3\chi_{29})\sin(3\theta))^2}$,
and $\theta_{TH}(\theta) = \text{atan}\left(\frac{3(\chi_{22}+\chi_{29})\sin(\theta)-(\chi_{22}-3\chi_{29})\sin(3\theta)}{3(\chi_{11}+\chi_{18})\cos(\theta)+(\chi_{11}-3\chi_{18})\cos(3\theta)}\right)$.

The dependence of the TH polarization angle $\theta_{TH}$ on the input linear polarization $\theta$ can be derived from Eq. (6). For the specific cases of $\theta = 0°$ and $\theta = 90°$, corresponding to horizontally- (H, along the *x* axis) and vertically- (V, along the *y* axis) polarized inputs, respectively, the TH light retains the same polarization as the fundamental, with $P_{TH}(0°) = \frac{1}{8}\varepsilon_0\chi_{11}A_0^3$ and $P_{TH}(90°) = \frac{1}{8}\varepsilon_0\chi_{22}A_0^3$.

The nonlinear polarization density characteristics arising from the utilization of circularly polarized inputs are instead described via Eqs. (3-4), by considering $\alpha = 0°$. The polarization response of the plain metasurface was quantified relative to the input polarization, to determine the nonlinear susceptibility tensor elements associated with the artificial THG process.

*Nonlinear polarization metagrating:* As a proof of concept, we studied a nonlinear polarization metagrating that consists of a periodic sequence of four cuboid meta-atoms along the *x*-axis, two featuring a rotation angle of $\alpha_0$ and the subsequent two of $-\alpha_0$, as shown in Fig. 1(d). Using Eq. (5), when a fundamental beam with H polarization ($\theta = 0°$) shines on the metastructure, the nonlinear polarization density amplitude can be written as $\tilde{P}_{3\omega,H_{in}} = \frac{A_0^3}{2} \begin{bmatrix} a_1 \cos 4\alpha + a_2 \cos 2\alpha + a_3 \cos 2\alpha + a_4 \\ -a_1 \sin 4\alpha + a_2 \sin 2\alpha - a_3 \sin 2\alpha \end{bmatrix}$. Here, the *x*-polarization component of the TH field remains unchanged upon flipping the sign of $\alpha$, resulting in its collinear emission with respect to the fundamental field. On the other hand, the *y*-polarized component undergoes a $\pi$-phase shift when the sign of $\alpha$ is changed, leading to diffraction of this TH field. Similarly, the nonlinear polarization density characteristics can be determined when a V-polarized input ($\theta = 90°$) is utilized, in this case revealing TH diffraction orders in the *x*-polarized component. Hence, such metagratings provide orthogonal linear polarizations between the zero- and other diffraction orders of the TH light.

*Nonlinear gradient metasurface:* We also expanded our theoretical estimations to a nonlinear gradient metasurface featuring meta-atoms with a linearly varied rotation angle along the *x*-axis (i.e., each column of the metasurface is characterized by meta-atoms progressively rotated along *x* by a fixed angle step, see Fig. 1(e)). This introduces a gradient control for the TH polarization density, allowing for the tailoring of the generated diffraction pattern using the angle step, $\alpha_1$, and the distance between neighboring meta-atoms. Such diffraction can be estimated with respect to the input polarization using the generalized Snell's law [42] combined with the theory described above. When RCP (LCP) fundamental light is used, the TH beam shows +1st and +2nd (-1st and -2nd) diffraction orders with polarizations RCP and LCP (LCP and RCP), respectively [23]. Beyond this, we show that the 0th order appears only when the fundamental light has a non-circular (i.e., linear or elliptical) polarization, a behavior different from that exhibited by linear optical gradient metasurfaces [43].

**Results**

We validated our metasurface design toolbox by fabricating and characterizing a-Si nonlinear metasurfaces featuring arrays of cuboid meta-atoms. As stated earlier, to simplify our analysis, we employed an off-resonant design at the fundamental operation wavelength $\lambda_f = 1596$ nm. To this end, the design parameters of the meta-atoms were determined through numerical simulations using Ansys Lumerical® (see details in **Methods**), including length $a = 420$ nm, width $b = 160$ nm, thickness $l = 425$ nm, and period (center-to-center distance between two neighboring meta-atoms) $D = 520$ nm (see inset of Fig. 2 (a)). The nanofabrication processes (including film deposition, patterning, lift-off, and etching) that were used to define the a-Si metasurfaces on a silica substrate are detailed in the **Methods** section.

We first characterized the plain metasurface consisting of meta-atoms with their long axis oriented along the *x*- axis, as shown in the scanning electron microscopy image in the inset of Fig. 2(a). The figure also presents the simulated and measured transmission characteristics of the device in the telecom wavelength range when linear polarizations, H and V, were used for illumination. The plain metasurface shows a high measured (simulated) transmissivity, $T_{exp} > 92\%$ ($T_{sim} > 94\%$), for both polarizations, ensuring an off-resonant operation at $\lambda_f \geq 1596$ nm. The simulated optical field distributions within the meta-atom under excitation of linearly polarized plane waves at $\lambda_f = 1596$ nm are presented in Section 2 of the **Supplementary Information**.

For the THG characterization of the fabricated metasurfaces, a tunable femtosecond laser was used as the fundamental light source. The emitted TH power was measured using a silicon-based calibrated photodiode paired with a lock-in amplifier. The setup also included an imaging module, comprising focusing lenses and a visible camera, to map the TH intensity profiles. Full details on the experimental setup can be found in **Methods** (see Section 3 of the **Supplementary Information** for a graphic sketch). The power dependence of the TH light emitted by the plain metasurface was measured by varying the fundamental peak intensity, as shown in Fig. 2(b), revealing the expected cubic trend. For ease of depiction, we only plotted the most efficient polarization, H, and one of the circular polarizations, RCP (an LCP input produced a similar TH response). TH images and spectra are shown in Section 4 of **Supplementary Information.**

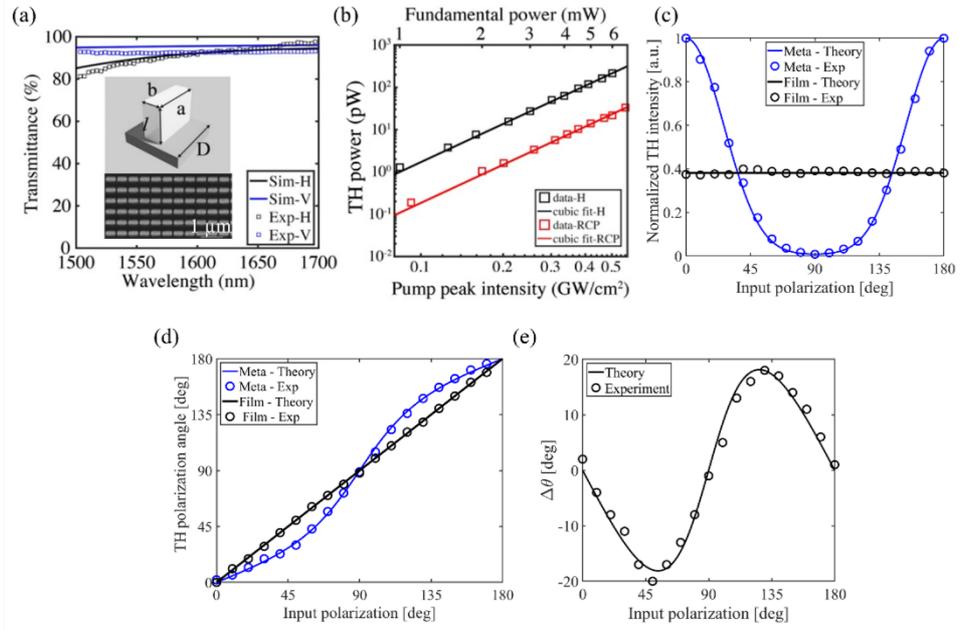

**Figure 2. Characterization of the plain metasurface.** (a) Simulated (solid curves) and measured (square markers) linear transmittance spectra when H (black) and V (blue) input polarizations were used for illumination. Inset: meta-atom sketch and scanning electron microscope image of the fabricated sample. Here, *a*, *b*, *l* and *D*, are the meta-atom length, width, thickness and period, respectively. (b) Power scaling measurement of the TH power (squares), when H (black) and RCP (red) input polarizations were used; solid lines show the corresponding cubic fit functions. (c) TH intensity (normalized to the metasurface peak value) measured as a function of the input linear polarization angle. The blue circles represent the experimental data of the plain metasurface, whereas the solid blue curve shows the expected theoretical trend retrieved from Eq. (6). (d) TH polarization angle measured as a function of the input polarization angle. The blue circles represent the metasurface experimental data, while the solid blue curve shows the metasurface data retrieved from Eq. (6). The black solid curves (theoretical) and circles (experimental) in (c,d) illustrate the trend followed by an a-Si film featuring the same thickness of the metasurface, where the TH intensity remains constant, and the TH polarization aligns with the pump polarization. (e) Theoretical (black solid curve) and experimental (black circles) difference in the TH polarization angle between the plain metasurface and the a-Si film.

Information on the nonlinear tensor elements is valuable in gaining a comprehensive understanding of nonlinear devices featuring artificial shape-induced spatial symmetries. We retrieved these effective nonlinear tensor elements by quantifying the TH conversion efficiencies of the fabricated plain metasurface relative to the polarization of the input and generated fields. As a result, the metasurface susceptibility tensor is estimated to be:

$$M_{meta} = \begin{pmatrix} \chi_{11} & 0 & \chi_{18} & 0 \\ 0 & \chi_{22} & 0 & \chi_{29} \end{pmatrix} =$$
$$\begin{pmatrix} 7.90 \times 10^{-18} & 0 & 4.67 \times 10^{-19} & 0 \\ 0 & 7.42 \times 10^{-19} & 0 & 1.38 \times 10^{-18} \end{pmatrix} (m^2/V^2).$$

Full details on the evaluation of the tensor elements can be found in Section 5 of the **Supplementary Information**. For the sake of comparison, we also measured the TH light from an a-Si film on the same substrate and featuring the same thickness of the metasurface, under identical illumination conditions, to showcase the geometry-enabled manipulation of the nonlinear response. The measured effective third-order susceptibility tensor of the film is given by:

$$M_{film} = \begin{pmatrix} \chi_{11} & 0 & 1/3\chi_{11} & 0 \\ 0 & \chi_{11} & 0 & 1/3\chi_{11} \end{pmatrix} =$$
$$\begin{pmatrix} 4.88 \times 10^{-18} & 0 & 1.63 \times 10^{-18} & 0 \\ 0 & 4.88 \times 10^{-18} & 0 & 1.63 \times 10^{-18} \end{pmatrix} (m^2/V^2).$$

As expected, due to the isotropic nature of the homogenous a-Si film, its $\chi_{11}$ and $\chi_{22}$ values are identical [40]. The isotropic film further features $\chi_{18} = \chi_{29} = \frac{1}{3}\chi_{11}$ [40], leading to the suppression of TH processes under circularly polarized fundamental inputs, as $\chi_{11} - 3\chi_{18} = \chi_{22} - 3\chi_{29} = 0$ (this can be inferred from Eq. (1)-(4), see Section 5 of the **Supplementary Information** for full details). In contrast, the plain metasurface shows a $\chi_{11}$ value that is an order of magnitude larger than $\chi_{22}$, as well as nondiagonal nonlinear tensor elements with distinct values, so that $\chi_{11} - 3\chi_{18} \neq \chi_{22} - 3\chi_{29} \neq 0$, which enables THG with circularly polarized input light.

We further analyzed the properties of the TH light generated by the plain metasurface when illuminated by a linearly polarized fundamental beam, by measuring its intensity and polarization as a function of the input polarization angle (from $\theta = 0°$ to $180°$). The TH intensity exhibits a significant variation relative to the input polarization state, as shown in Fig. 2(c) (blue circles), following closely the theoretical prediction (blue line). This behavior arises from the strong modulation of the nonlinearity introduced by the cuboid shape of the meta-atoms, which results in geometry-dependent nonlinear susceptibility tensor terms. In contrast, the TH intensity of an a-Si film is constant with varying input polarization angle (black circles), as also predicted theoretically (black line). The TH polarization measurements were performed by incorporating a linear polarizer (polarization analyzer) prior to detection, with the results shown in Fig. 2(d). The reported TH polarization angle corresponds to the analyzer angle for which TH transmission was maximized (for an orthogonal analyzer angle, TH was strongly attenuated with a contrast of 1/100). Notably, the significant deviation of the TH polarization angle relative to the input polarization is evident here (blue circles: experiment; blue line: theory), in contrast to the isotropic film, where the TH polarization state matches the input polarization (black line). To further highlight this deviation, the theoretical (black solid curve) and measured (black circles) TH polarization angle difference between the plain metasurface and the a-Si film is presented in Fig. 2(e), showcasing values ranging up to almost $\pm 20°$ for certain input polarizations. All the theoretical curves were evaluated using Eq. (6) together with the retrieved susceptibility tensor values. Overall, these results underline the high potential of subwavelength patterning in altering the intrinsic nonlinear characteristics of a material, thus paving the way for the rational design of novel functional nonlinear devices.

As a first example, we designed a nonlinear polarization metagrating (see Fig. 3(a)), as detailed in the 'Device design' section above. Here, the grating period ($D'$) along the *x*-direction consists of four meta-atoms, each separated by a distance $D = 520$ nm, with two featuring a rotation angle of $\alpha = 22.5°$ and the subsequent two of $\alpha = -22.5°$, resulting in $D' = 4D = 2.08$ µm. The measured spatial profiles of the TH intensity from the metagrating for various input polarizations are shown in Fig. 3(b), together with the numerical simulation outcomes. Regarding the latter, the spatial TH field distribution at the image plane was simulated by integrating the emitted light field from the meta-atoms located at the metasurface plane. The experimentally measured deflection angles of the $\pm 1^{st}$ diffracted beams were found to be ~$\pm 17°$, in good agreement with the expected values of $\pm 15°$, the slight difference being attributed to measurement conditions and/or fabrication discrepancies. The complete characterization of the TH spectra and power scaling measurements of the diffraction orders are included in Section 6 of the **Supplementary Information**. The nonlinear metagrating was further studied by injecting H- and V- polarized fundamental light while characterizing the emitted TH polarization using a linear polarizer as a polarization analyzer. It was observed that, when an H- (or V-) polarized fundamental beam was used for illumination, the zero-order TH output featured the same polarization, i.e., H (or V), while the first diffraction orders showed the orthogonal polarization, i.e., V (or H) (see Fig. 3(c) and (d)). The experimental results agree well with the model prediction discussed in the 'Device Design' section. Such a metagrating is thus capable of generating TH polarization states that are orthogonal to and spatially separated from the fundamental light, an interesting feature for potentially advancing polarization-based imaging technologies [30, 33] and sensing applications [44].

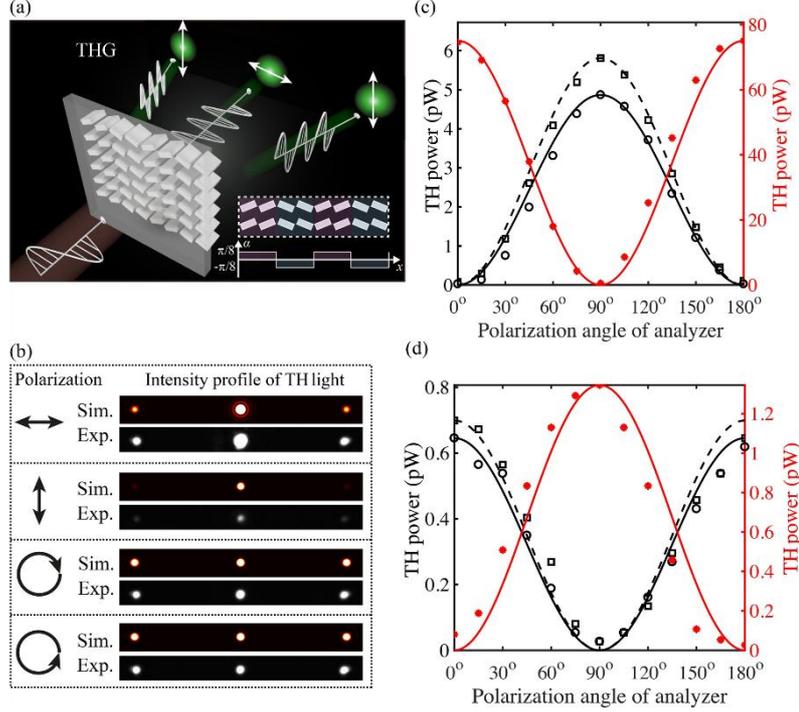

**Figure 3. Characterization of a phase-grating metasurface (metagrating).** (a) Sketch of the nonlinear metagrating working principle. Inset: illustration of the metasurface design. Each grating period consists of four blocks with spatially varying rotation angles of $\pm 22.5°$ along the *x*-direction. The grating period is 2.08 μm with duty cycle of 50%. (b) Simulated and measured TH intensity profiles of the diffracted light, showcasing $0^{th}$ and $\pm 1^{st}$ orders, when different fundamental polarizations (H, V, RCP, and LCP) were incident onto the metagrating. Note that both the simulated and experimentally captured images were obtained under saturated conditions to better reveal the diffraction orders. (c) and (d) show the measured TH powers of $0^{th}$ (red dots), $+1^{st}$ (black circles), and $-1^{st}$ (black squares) diffraction orders as a function of the polarization angle of the analyzer under the illumination of H- (c) and V- polarized (d) fundamental light, respectively. Black and red solid curves represent the theoretical estimations, with their peak values normalized to the maxima of the measured data.

Finally, to further support our theoretical and device engineering framework, a nonlinear gradient metasurface (see Fig. 4(a)), comprising meta-atoms that are gradually rotated along the *x*-axis with an angle step $\alpha_1 = 22.5°$, was developed and investigated. The simulated and measured diffraction patterns of the TH light are shown in Fig. 4(b), when linear and circular polarization inputs were used. The $1^{st}$ and $2^{nd}$ diffraction orders angles were measured at 8.9° and 17.8°, respectively, showing slightly larger values than the expected simulation outcomes of 7.2° and 14.5°. We characterized the generated ±1st diffraction orders by adjusting the ellipticity of the fundamental light. More in detail, we tuned the fundamental polarization from LCP to RCP by employing a half waveplate (HWP) followed by a quarter waveplate (QWP) (note that the initial polarization of the pump light was horizontal). The QWP angle was set to 0º (with respect to the *x*-axis), whereas the HWP was rotated so that a range of input polarizations was obtained, i.e., in sequence: left-handed circular and elliptical polarizations, linear polarization, and finally right-handed elliptical and circular polarizations. As shown in Fig. 4(c), the power of the +1$^{st}$ TH diffraction order initially shows a plateau, due to the appearance of the third term in Eq. (1), and then continuously decreases to zero. The power of the -1$^{st}$ order follows a similar, but opposite, trend. This behavior is also observed in the theoretically estimated values obtained using Eq. (1) (solid curves in Fig. 4(c)), which highlights the ability of the developed design toolbox to accurately predict the polarization response of nonlinear metadevices.

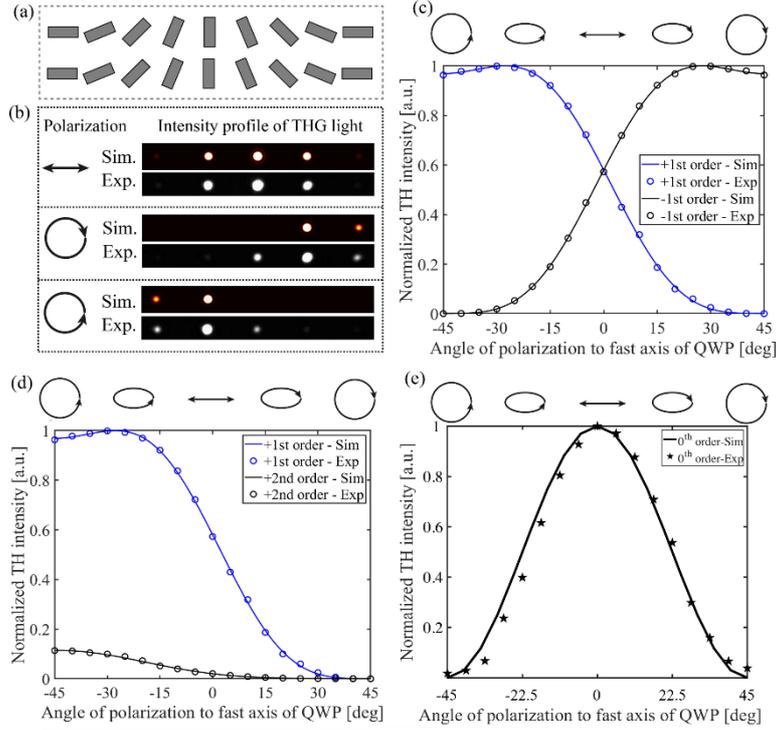

**Figure 4. Characterization of a phase-gradient metasurface.** (a) Illustration of the metasurface design. The meta-atoms are gradually rotated along the x-direction with an angle step of 22.5°. (b) Simulated and measured intensity profiles of the diffracted TH light, exhibiting 0$^{th}$, ±1$^{st}$, and ±2$^{nd}$ orders, for different fundamental polarization inputs. (c) Theoretical estimations (solid curves) and experiment results (circle markers) of TH intensity with respect to the ellipticity of the input fundamental light. The blue and black colors represent the +1$^{st}$ and -1$^{st}$ orders, respectively. (d) Theoretical estimations (solid curves) and experiment results (circle markers) of TH intensity of the +1$^{st}$ (blue color) and +2$^{nd}$ (black color) diffraction orders as a function of the fundamental light polarization. The simulated and measured power of the +1$^{st}$ and +2$^{nd}$ orders are normalized to the maximum value of the simulated and measured +1$^{st}$ order, respectively. (e) The normalized power trend of the 0$^{th}$ order with respect to the fundamental light polarization. The solid curve represents the simulation, while the star markers are the measured data.

Our model also allows to quantitively compare the TH intensities of the 1$^{st}$ and 2$^{nd}$ diffraction orders. Note that the TH ±1$^{st}$ orders of diffraction feature the same polarization handedness of the fundamental light for input circular polarization, whereas the ±2$^{nd}$ orders exhibit the opposite handedness (see Section 7 of the **Supplementary Information** for more details). Fig. 4(d) shows the theoretically estimated and experimentally retrieved intensities of the +1$^{st}$ and +2$^{nd}$ TH orders as a function of the fundamental light ellipticity. The measured data exhibits the expected trend as a function of the input polarization, with an excellent agreement with theory for what concerns the ratio between the TH power emitted in the two diffraction orders. Furthermore, as predicted by our model, when the polarization of the fundamental light is changed from purely circular to elliptical or linear, which are superpositions of both LCP and RCP components, the fourth term in Eq. (1) emerges, resulting in the appearance of the $0^{th}$ diffraction order. Fig. 4(e) depicts the simulated and measured power trend of the $0^{th}$ order of the TH light with respect to the ellipticity of the fundamental light, demonstrating a consistent behavior between the two outcomes. This type of device architecture shows promise for generating complex optical states through nonlinear wave-mixing processes, by accessing the spatial (via multiple diffraction orders), spectral (because of the broadband frequency response, unrestricted by phase-matching rules), and polarization degrees of freedom, applicable to high-density information encoding schemes [45-47] potentially extending to quantum communication applications [48, 49].

**Discussion and Conclusion**
We have presented a comprehensive mathematical description for the THG process occurring in dielectric meta-atoms endowed with artificial optical nonlinearities originating from their geometrical features and symmetry. The polarization properties of the TH light have been fully described in relation to the fundamental light polarization, complementing the theory already developed for nonlinear optical processes in metasurfaces [23]. The proposed model provides a convenient way to design unique nonlinear metasurface devices, by employing both the geometrical parameters (i.e., the characteristic dimensions) and the spatial symmetries of the meta-atoms for the realization of a novel, compact, and flexible framework for nonlinear photonic manipulations.

To show the effectiveness of our metasurface design toolbox, we have first studied a plain metasurface consisting of an array of a-Si cuboid meta-atoms and retrieved the artificial third-order susceptibility tensor matrix associated with this geometry. Building on this characterization, nonlinear metagrating and nonlinear gradient metasurface devices have been investigated for the generation of tailorable polarization properties across various diffraction orders. Beyond possible applications in nonlinear polarization imaging and in the generation of complex vector beams [33, 34, 36], the presented nonlinear metasurfaces also show the potential to be adapted towards the development of quantum light sources with well-defined polarization states, emitted in, e.g., the first diffraction order, while providing effective suppression of the classical pump field transmitted in the $0^{th}$ order. As is well-known, the co-propagation of classical and non-classical fields introduces unavoidable noise during the detection of quantum states [50], which is a significant drawback for practical implementations. Our study suggests that precisely shaped nonlinear metadevices could effectively address this challenge by separating classical–noise from quantum fields both spatially and in terms of polarization, which may find interesting applications in nonlinear metasurface-based quantum schemes, e.g., for imaging and sensing [51, 52].

The presented theory can be readily adapted to different meta-atom symmetries, for example, cross-shaped structures corresponding to the tetragonal crystal class, i.e., where, $\chi_{11} = \chi_{22}$, and $\chi_{18} = \chi_{29}$ [40]. In such case, the second and third terms in Eq. (1) vanish since $a_2 = a_3 = 0$. Hence, under the illumination of circularly polarized fundamental light, only the TH light with the opposite circular polarization can be generated, matching the conclusions of Refs. [23, 29]. While, for simplicity, non-resonant meta-atoms were considered in this work, the extension of the model to resonant structures, which can enhance the nonlinear conversion efficiency [53, 54], is certainly of interest for future investigations. The theoretical description can also be adjusted for other frequency conversion processes, such as second-order nonlinear interactions, e.g., second-harmonic, sum- and difference- frequency generation, as well as for the investigation of the properties of light emitted from asymmetric meta-atoms [55, 56].

## Methods
### Electromagnetic simulations
For the linear simulations of the investigated metasurfaces (i.e., transmission profile and electric field distribution), we employed the commercial software Ansys Lumerical®. For the recording of the linear transmission profile (Fig. 2(a)), a plane wave was used for excitation, and the transmitted light was collected in the far field through a monitor. To retrieve the electric field distribution of the fundamental light within the meta-atoms, a monitor was placed at the nanostructure location for the design wavelength of $\lambda_f = 1596$ nm. Such results are shown in Section 2 of the **Supplementary Information**. For all the simulations, periodic boundary conditions were employed at the metasurface plane, while perfectly-matched layer boundary conditions were used above and below this plane. The dispersion of the complex refractive index of both amorphous silicon and silica was properly accounted for, so to obtain accurate information about the metasurface behavior.

### Sample fabrication
A 425 nm-thick a-Si film was grown on a fused silica substrate using Plasma Enhanced Chemical Vapor Deposition at 350°C, with a high silane flux diluted in argon. An electron beam (e-beam) sensitive resist, polymethyl methacrylate (PMMA) was uniformly spin-coated onto the grown film at 2000 rpm, with an approximate thickness of 50 nm. The sample was then post-baked at 180°C for 5 minutes. To prevent the sample from charging during the e-beam exposure, a thin conductive polymer layer was coated onto the PMMA. The sample was then exposed using an e-beam lithography system, Vistec VB300. After development, a 10 nm-thick chromium layer was uniformly deposited, which acted as a hard mask for the etching process. The lift-off was then performed in Remover PG at 80°C, leading to an exposed silicon film and the chromium mask. An etchant of $SF_6:C_4F_8$ was used in an Oxford PlasmaLab 150 ICP to etch the exposed silicon completely. The concentration of the etchant was tailored to achieve vertical sidewalls. Lastly, the chromium mask was removed in an acid solution, thus resulting in the final a-Si metasurface device on a fused silica substrate.

### Optical characterization
A tunable femtosecond laser system (*Pharos* Yb:KGW regenerative amplifier equipped with an optical parametric amplifier *Orpheus* from *Light Conversion*) featuring a pulse duration of $\tau \sim 150$ fs and a repetition rate of 500 kHz was used to characterize the metasurface samples. The full-width-at-half-maximum spectral bandwidth at $\lambda_f = 1596$ nm was estimated to be ~31 nm. A broadband linear polarizer, an achromatic half-wave plate, and a zero-order quarter-wave plate were used to control the polarization of the fundamental light (note that the quarter waveplate was removed when generating the linear polarization states). A telescope configuration consisting of two $CaF_2$ lenses with focal lengths of 250 mm and 20 mm was used to reduce the input fundamental beam waist down to 130 µm (half-width at $1/e^2$ of the maximum intensity) at the sample position. An objective lens (NA = 0.4) was employed to collect the TH emitted radiation, followed by two short-pass filters to suppress the fundamental light. Diffraction orders emitted by the metasurfaces were selected using a slit mounted on a translation stage. The TH power was measured using a femtowatt silicon-based calibrated photodiode (Newport New Focus 2151) paired with a lock-in amplifier synchronized to a 500 Hz chopper frequency. The employed photodiode can measure powers down to tens of fW. A schematic of the experimental setup is shown in Section 3 of the **Supplementary Information**.

**Acknowledgments**
Work at INRS was supported by the Natural Sciences and Engineering Research Council of Canada (NSERC) through Discovery grant and Alliance Consortia Quantum programs, as well as by the Canada Research Chair Program. L. R. and R. M. acknowledge further support from the Fonds de Recherche du Québec – Nature et Technologies (FRQNT, Projet de Recherche en Équipe, DOI: https://doi.org/10.69777/282002), and Photonique Quantique Québec (PQ2). Work at the Molecular Foundry was supported by the Office of Science, Office of Basic Energy Sciences, of the U.S. Department of Energy under Contract No. DE-AC02-05CH11231. The Molecular Foundry thanks S. Dhuey for assistance with the electron-beam lithography. F. Y. acknowledges support from FRQNT (postdoctoral research scholarship (B3X)). N. M. acknowledges funding from the Mitacs Elevate Thematic Quantum Science program.